\documentclass[a4paper,11pt]{article}
\pdfoutput=1 

\usepackage{jheppub} 

\usepackage[T1]{fontenc} 
\usepackage{mathrsfs}
\usepackage{latexsym,bm}
\usepackage{graphicx}
\usepackage{indentfirst}
\usepackage{amsmath}
\usepackage{amssymb}
\usepackage{color}
\usepackage{hyperref}
\usepackage{epsfig}
\usepackage[titletoc]{appendix}
\usepackage{multirow}
\usepackage{rotating}
\usepackage{epstopdf}
\usepackage{extarrows}

\newcommand{\bea}{\begin{eqnarray}}
\newcommand{\eea}{\end{eqnarray}}
\newcommand{\no}{\nonumber\\}
\newcommand{\order}[1]{\mathcal{O}\left(#1\right)}
\newcommand{\sulr}{{SU$(N)_L\times$SU$(N)_R$}}
\newcommand{\suv}{{SU$(N)_V$}}
\newcommand{\tr}[1]{\left\langle#1\right\rangle}

\title{\boldmath Subtraction of power counting breaking terms in chiral perturbation theory:
spinless matter fields}

\author[a]{Meng-Lin Du,}
\author[b]{Feng-Kun Guo}
\author[a,c]{and Ulf-G.~Mei{\ss}ner}

\affiliation[a]{Helmholtz-Institut f\"ur Strahlen- und Kernphysik and Bethe Center for Theoretical Physics,\\
Universit\"at Bonn,\\
Nu{\ss}allee 14-16,  D--53115 Bonn, Germany}
\affiliation[b]{CAS Key Laboratory of Theoretical Physics, Institute of
Theoretical Physics,\\
Chinese Academy of Sciences,\\
Zhong Guan Cun East Street 55, Beijing 100190, China}
\affiliation[c]{Institute for Advanced Simulation, Institut f{\"u}r
Kernphysik and J\"ulich Center for Hadron Physics,\\
Forschungszentrum J{\"u}lich, \\
Wilhelm-Johnen-Stra{\ss}e, D-52425 J{\"u}lich, Germany}

\emailAdd{du@hiskp.uni-bonn.de}
\emailAdd{fkguo@itp.ac.cn}
\emailAdd{meissner@hiskp.uni-bonn.de}

\abstract{When matter fields are included in chiral perturbation theory, the nonvanishing
mass in the chiral limit introduces a new energy scale so that the loop diagrams
including such matter field propagators spoil the usual  power counting.
However, the power counting breaking terms can be absorbed into counterterms in
the chiral Lagrangian. In this paper, we systematically derive these terms to
 leading one-loop order (next-to-next-to leading order in the chiral
expansion) at once by calculating the generating functional using the path integral.
They are then absorbed by counterterms in the next-to-leading order Lagrangian.
The method can be extended to calculating power counting breaking terms for
other matter fields.}

\begin{document}
\maketitle
\flushbottom

\section{Introduction}

Quantum chromodynamics (QCD) is the fundamental theory for the strong
interactions of quarks and gluons. An important feature of QCD is asymptotic
freedom,  which implies that in the high energy region the observables can be
expanded in powers of the strong coupling constant $\alpha_s$.
However, at low energies, the theory is highly non-perturbative since $\alpha_s$
becomes large. The failure of a perturbative expansion in $\alpha_s$ requires
alternative approaches to investigate the low-energy behavior of the strong
interactions. Chiral perturbation theory
(ChPT)~\cite{Weinberg:1978kz,Gasser:1984gg} as the low-energy effective theory
of QCD (and the Standard Model) presents a very important tool in this field. It
is based on the spontaneous and explicit breaking of chiral symmetry. The
massless QCD Lagrangian has a global symmetry U$(N)_L\times$U$(N)_R$ at the
classical level, with $N$ the number of the light quark flavors under
consideration.
However, the singlet axial current develops an  U$(1)_A$ anomaly at the quantum
level~\cite{Adler:1969gk,Christos:1984tu}. In other words, the QCD Lagrangian
has a chiral symmetry $\text{SU}(N)_L\times\text{SU}(N)_R\times\text{U}(1)_V$ in
the limit of vanishing quark masses. While the U$(1)_V$ symmetry is manifest as
baryon number conservation, the chiral symmetry \sulr~ is spontaneously broken
into its vectorial subgroup SU$(N)_V$, and $N^2-1$ Goldstone bosons emerge.
Meanwhile, the nonvanishing quark masses break the chiral symmetry explicitly,
giving a finite but small mass to the Goldstone bosons.
The spontaneously and explicitly broken chiral symmetry plays a central role in
the low-energy domain of the strong interactions, and lays the basis for ChPT
which is a model-independent effective field theory equivalent to QCD at low
energies.

As an effective theory, ChPT provides a powerful and successful framework for
studying hadron physics at low energies.
It explores a scale separation. The hard energy scale is given by the scale of
chiral symmetry breaking $\Lambda_\chi\sim 1~\text{GeV}$, which corresponds to
the occurrence of ``typical" hadrons. The soft energy scale is provided by the
low-momenta of Goldstone bosons and external fields as well as the Goldstone
boson masses (or equivalently the light quark masses). The
Goldstone bosons as the low-energy degrees of freedom are treated explicitly,
while the high-energy modes are integrated out and the information is encoded
into the coupling constants, the so-called low-energy constants (LECs).
Due to the small values of the light quark ($u$, $d$ and $s$) masses,  the
explicit chiral symmetry breaking effects can be treated perturbatively, and one
can construct the chiral Lagrangian order by order in a double expansion in the
small momenta, denoted as $p$, of the Goldstone bosons and the light quark
masses $m_q=\order{p^2}$, where the power of $p$ is called the chiral dimension.
Since the effective Lagrangian contains an
infinite number of local operators, a proper power counting rule is necessary to
determine the relative importance of the various terms and the Feynman diagrams
based on them. The power counting~\cite{Weinberg:1978kz} for a given
term and Feynman diagram may be obtained by rescaling the external momenta $p\to
t p$ and the light quark masses $m_q \to t^2 m_q$ and counting the power of $t$,
which gives the chiral dimension.
Applying this  power counting scheme, a relation between the momentum
expansion and the loop expansion is well established for the purely Goldstone
boson sector, i.e. when there are no matter fields.

The systematic studies of the purely Goldstone boson ChPT up to one loop, i.e.
$\order{p^4}$, have been performed by Gasser and Leutwyler in their classical
papers~\cite{Gasser:1983yg,Gasser:1984gg}. The full generating functional
(including one-loop graphs) at $\order{\phi^4}$, i.e. with at most four external
Goldstone bosons, was obtained up to $\order{p^4}$, which suffices to calculate
all relevant two-point correlation functions. The one-loop diagrams with all
vertices from the $\order{p^2}$ Lagrangian appear at the order $\order{p^4}$. They
are ultraviolet (UV) divergent and the UV divergences can be renormalized by the
counterterms in the $\order{p^4}$ Lagrangian using dimensional regularization
which preserves all the symmetries under consideration.

Matter fields which have nonvanishing masses in the chiral limit can be included
in ChPT as well. ChPT including light baryons is known as baryon ChPT. There
exists a notable power counting breaking (PCB) issue in baryon
ChPT~\cite{Gasser:1987rb}:
the naive power counting rule does not work for loop diagrams containing baryon
propagators and all such diagrams start contributing at $\order{p^2}$ using the
$\overline{\text{MS}}$ subtraction scheme of  dimensional
regularization.~\footnote{The closed matter field loops are not taken into
account because they are real below the two matter-field threshold and are
counted as $\order{1}$.
Their contributions can be absorbed by a redefinition of
LECs.~\cite{Gasser:1987rb,Roessl:1999iu}} This problem is caused by the
nonvanishing mass of matter field in the chiral limit, which then presents an
energy scale of $\order{1}$. Various approaches were proposed to address the PCB
issue, e.g.
the heavy baryon ChPT (HBChPT)~\cite{Jenkins:1990jv,Bernard:1992qa}, the
infrared regularization (IR)~\cite{Becher:1999he}, and the
extended-on-mass-shell (EOMS) scheme~\cite{Fuchs:2003qc}. Likewise, ChPT
including other matter fields has the same PCB problem. Note further that  the approaches used
for baryons can be employed to ChPT including other matter fields.

In HBChPT, the heavy components of the baryon fields are integrated out. The
dependence of the matter field mass $m$ is then removed from propagators by
expanding the Lagrangian in powers of $1/m$. The resulting loops satisfy the
power counting. However, the $1/m$ expansion sometimes produces incorrect
low-energy analytic properties~\cite{Bernard:1995dp}. The expansion series near
the anomalous threshold fails to converge and an infinite number of terms are
needed.\footnote{Note, however, that this can be overcome by using the extended
propagator $i/(v\cdot k + k^2/2m)$ instead of the strict HB propagator $i/v\cdot
k$, see e.g. Ref.~\cite{Bernard:1993ry}.} A manifestly Lorentz covariant
regularization scheme which preserves the analytic structure and the power
counting  can also be formulated. A very important step was made by Ellis and
Tang~\cite{Tang:1996ca,Ellis:1997kc}.
They noted that the soft-momentum part of a loop diagram is infrared singular
and the PCB terms, coming from the hard-momentum modes only, are a local
polynomial in small momenta and Goldstone boson masses and can be absorbed into
the LECs of the most general chiral Lagrangian.
Based on their work, Becher and Leutwyler proposed the IR scheme which isolates
the infrared singular parts of the loops by extending the Feynman parameter
integration upper bound from unity to infinity~\cite{Becher:1999he}.
Due to the fact that the infrared regular parts of loops can be obtained by
expanding the integrand in small quantities and then integrating each
term~\cite{Gegelia:1994zz}, the EOMS scheme was proposed, in which additional
subtractions beyond the $\overline{\rm MS}$ scheme are performed to get rid of
the PCB terms.

In this paper, we will study the one-loop generating functional of correlation
functions with up to four external particles for the chiral Lagrangian with
spinless matter fields in the fundamental representation of $\text{SU}(N)$. Such
a theory can be applied to study kaon-pion scattering by treating kaons as
matter fields\cite{Roessl:1999iu}, and it is expected to have  a better
convergence of the chiral expansion than that of the normal $\text{SU}(3)$ ChPT
which treats kaons as Goldstone bosons as well. It can also be employed to
investigate the interaction between heavy mesons and Goldstone
bosons~\cite{Guo:2009ct,Yao:2015qia}.
A systematic one-loop renormalization of this theory was done in
Ref.~\cite{Du:2016ntw} by calculating the divergence of the effective action
using the background field method and heat kernel technique. This paper presents
a step further as we will provide a systematic subtraction of the PCB terms at
the Lagrangian level at the leading one-loop order, i.e. $\order{p^3}$, for such
a theory. All PCB terms will be obtained at once by calculating the
corresponding part in the one-loop generating functional.

This paper is organized as follows. In Section~\ref{seclag}, the power counting
formula and its breaking in the presence of matter fields are briefly introduced;
the relevant notation and the chiral Lagrangian for spinless matter fields in
the $\text{SU}(N)$ fundamental representation are given to the
next-to-next-to-leading order (NNLO).
The explicit one-loop generating functional is derived in Section~\ref{secfunc}.
In Section~\ref{secpcb}, the PCB terms are obtained and subtracted in the EOMS
scheme by redefining the $\order{p^2}$ LECs.
Section~\ref{summary} presents a brief summary.

\section{Power counting and effective Lagrangian}
\label{seclag}

We denote the matter fields and Goldstone boson fields as $P$ and $\phi$,
respectively.
To introduce the effective Lagrangian, one has to specify the power counting
rules. At low energies, the momenta, as well as the masses $M_\phi$, of
Goldstone bosons are counted as $\order{p}$. However, the nonvanishing mass of
the matter field, $m_P$, in the chiral limit introduces a new energy scale. Since
matter fields are normal hadrons which are not Goldstone bosons, their masses are of
the order of $\Lambda_\chi$ and should be counted as $\order{1}$. The temporal
component of the momentum of matter field should then be counted in the same
way. Yet, at low energies when the three-momentum of matter field is small, one
may count $q^2-m_P^2$ as $\order{p}$, and thus the propagator as
$\order{p^{-1}}$. The Goldstone boson
propagator $i/(q^2-M_\phi^2)$ is counted as $\order{p^{-2}}$. One can then
derive the chiral dimension $n$ for a given Feynman diagram as
\bea
n=4L+\sum_k V_k-2I_\phi-I_P, \label{pcr}
\eea
where $L$, $V_k$, $I_\phi$ and $I_P$ denote the numbers of loops, the $k^{\rm th}$
order vertices, internal Goldstone boson propagators and internal matter field
propagators, respectively.

For a specific Feynman graph, if there exist terms whose chiral order is lower than
that given by Eq.~\eqref{pcr}, those terms are called PCB terms. The power
counting given in Eq.~\eqref{pcr} works well and no PCB term exists for purely
Goldstone boson ChPT. However, PCB terms show up when matter field propagators
enter the loop integrals which are calculated using dimensional regularization
with the $\overline{\rm MS}$ scheme. This is due to the existence of the new
energy scale $m_P$. It is worth noting that the matter field mass in the chiral
limit, denoted as $m$, is of the same chiral order as the physical masses.
In the EOMS scheme, these PCB terms are absorbed into the redefinition of the LECs
so as to make the amplitudes have a power counting consistent with
Eq.~\eqref{pcr}.

We only consider the case involving a single matter field. The relevant
generating functional is defined as
\bea
e^{iZ[j,J,J^\dag]}=\langle 0~\text{out}|0~\text{in}\rangle_{j,J,J^\dag}=\int
[d\phi][dPdP^\dag]\exp \Big\{ i\int d^4x \left[\mathcal{L}_\phi (j) +
\mathcal{L}_{\phi P}(J,J^\dag) \right]\Big\}, \label{genfunc}
\eea
where $\mathcal{L}_\phi$ and $\mathcal{L}_{\phi P}$ denote the purely Goldstone
boson effective Lagrangian and the Lagrangian for the interaction between
Goldstone bosons and a single matter field, respectively. Further,  $J$ and $J^\dag$
denote the external sources coupled to the matter fields, and $j$ collects
various external fields coupled to the Goldstone bosons, i.e. vector $v_\mu$,
axial-vector $a_\mu$, scalar $s$ and pseudoscalar $p$. As usual, the explicit
$\text{SU}(N)$ symmetry breaking effect by quark masses $\mathcal{M}$ will be
included in the Lagrangian through the external scalar source $s$ via  $\chi=2B_0
s$, with $s=\mathcal{M}=\text{diag}(m_u,m_d,\ldots)$, where $B_0$ is a constant
related to the quark condensate. The effective Lagrangians can be expanded in a
power series as $$\mathcal{L}^{}_\phi =\sum_{n=1}^{\infty}\mathcal{L}_\phi^{(2n)},
\qquad \mathcal{L}^{}_{\phi P}=\sum_{n=1}^{\infty}\mathcal{L}_{\phi P}^{(n)},$$
where the upper indices indicate the chiral dimensions.

To construct the effective Lagrangians respecting the chiral symmetry, we collect
the Goldstone bosons in a $N\times N$ unitary matrix $U(x)$,
\bea
U(x)=u^2(x)=\exp{\left( \frac{i\phi}{F_0}\right)},
\eea
where $F_0$ is the pion decay constant in the chiral limit and $\phi$
is expanded in the $N^2-1$ traceless Hermitian basis $\phi=\lambda^a\phi^a$,
with $\lambda^a$ and $\phi^a$ the $\text{SU}(N)$ generators and the
Goldstone boson fields, respectively. $U(x)$ and $u(x)$ transform under \sulr~as
\bea
U\mapsto g_R U g_L^\dag, \qquad u\mapsto \sqrt{g_R U g_L^\dag}\equiv g_R u K^\dag =K u g_L^\dag,
\eea
where $g_L \in \text{SU}(N)_L$, $g_R \in \text{SU}(N)_R$, and   $K$ is a
nonlinear function of $g_L$, $g_R$ and $U(x)$. The compensator
field $K$ represents an element of the subgroup SU$(N)_V$, and it reduces to
$K=g_L=g_R$, independent of $U(x)$, for a SU$(N)_V$ transformation, i.e. when
 $g_L=g_R$ (see, e.g., Ref.~\cite{Ecker:1994gg}). It is
convenient for building up effective Lagrangians respecting the symmetry
constraints to construct the matter fields so that they transform under \sulr~as
\bea
P\mapsto P K^\dag\,, \qquad P^\dag \mapsto K P^\dag\,.
\eea
Here, we have followed the convention of Ref.~\cite{Du:2016ntw} to define
$P^\dag$ and $P$ in the fundamental and anti-fundamental representation of
SU($N$), respectively.

For the convenience of constructing a chirally invariant Lagrangian, we
introduce chiral-covariant derivatives as
\bea
D_\mu P^\dag =\partial_\mu P^\dag + \Gamma_\mu P^\dag, \qquad  D_\mu
P=\partial_\mu P +P \Gamma^\dag_\mu\,,
\eea
with the chiral connection
$\Gamma_\mu=\frac{1}{2}\left[u^\dag(\partial_\mu-ir_\mu )u+u(\partial_\mu
-il_\mu )u^\dag\right]$, and further $r_\mu=v_\mu+a_\mu$ and
$l_\mu=v_\mu-a_\mu$.
One can construct three other building blocks
\bea
u_\mu=i\left[u^\dag(\partial_\mu-ir_\mu )u+u(\partial_\mu -il_\mu
)u^\dag\right],\quad \chi_{\pm}=u^\dag\chi u^\dag\pm u\chi^\dag u,
\eea
as well as the covariant derivatives on them such as $\nabla_\mu
u_\nu=\partial_\mu u_\nu+[\Gamma_\mu,u_\nu]$.  These operators
transform under \sulr~as
\bea
u_\mu \mapsto K u_\mu K^\dag\,, \quad \chi_{\pm}\mapsto K \chi_{\pm} K^\dag\,,
\quad \nabla_\mu u_\nu \mapsto K \nabla_\mu u_\nu K^\dag\, .
\eea

The power counting rules for these building blocks are
\bea
  D_\mu P^{(\dag)} \sim \order{1}, \quad & D_\mu D_\nu P^{(\dag)} \sim \order{1},
  \quad & (D_\mu D^\mu+m^2)P^{(\dag)} \sim \order{p}, \no u_\mu \sim
  \order{p},\quad & \chi_{\pm}\sim \order{p^2},\quad &  \nabla_\mu u_\nu
  \sim \order{p^2}.
\eea
Based on the power counting for the Feynman graphs, the calculation of relevant
physical observables up to $\order{p^3}$ requires the effective chiral Lagrangian
\bea
\mathcal{L}_3^\text{eff}=\mathcal{L}_{\phi P}^{(1)}+\mathcal{L}_{\phi P}^{(2)}+\mathcal{L}_{\phi P}^{(3)}+\mathcal{L}_{\phi}^{(2)}+\mathcal{L}_{\phi}^{(4)}.
\eea
The Lagrangians can be found in
Refs.~\cite{Gasser:1984gg,Guo:2009ct,Yao:2015qia,Du:2016ntw,Bijnens:2006zp}. For
completeness, we list the relevant terms here. The Lagrangian for
the matter fields is
\bea
\mathcal{L}_{\phi P}^{(1)} & = &  D_\mu P D^\mu P^\dagger -m^2 PP^\dagger, \no
\mathcal{L}^{(2)}_{\phi P} & = & P\left[-h_0\langle\chi_+\rangle-h_1{\chi}_+
+ h_2\langle u_\mu u^\mu\rangle-h_3u_\mu u^\mu\right] {P}^\dag \nonumber\\
 &   &  + D_\mu P\left[{h_4}\langle u_\mu
u^\nu\rangle-{h_5}\{u^\mu,u^\nu\}\right] D_\nu {P}^\dag\ , \nonumber\\
\mathcal{L}^{(3)}_{\phi P} & = & \bigg[ i~g_1 P[\chi_-,u_\nu]D^\nu
P^\dagger+g_2 P[u^\mu,\nabla_\mu u_\nu+\nabla_\nu u_\mu]D^\nu
P^\dagger  \nonumber\\
 &   &  +g_3 P\left[u_\mu,\nabla_\nu u_\rho\right]D^{\mu\nu\rho}P^\dagger+ g_4
P \nabla_\nu \chi_+ D^\nu P^\dagger +g_5 P\langle \nabla_\nu \chi_+\rangle D^\nu
P^\dagger+ h.c.\bigg] \nonumber\\
 &   &  + i~\gamma_1 D^\mu P f_{\mu\nu}^+ D^\nu P^\dagger+  \gamma_2
P[u^\mu,f^-_{\mu\nu}]D^\nu P^\dagger~, \label{lagphiP}
\eea
and the Goldstone boson  Lagrangians has the form
\bea
\mathcal{L}_\phi^{(2)}  & = &  \frac{F_0^2}{4}\langle u_\mu u^\mu \rangle+
\frac{F_0^2}{4}\langle\chi_+\rangle, \no \mathcal{L}_\phi^{(4)}
& = & L_0\langle u^\mu u^\nu u_\mu u_\nu\rangle + L_1 \langle u_\mu u^\mu
\rangle^2+ L_2 \langle u^\mu u^\nu\rangle\langle u_\mu u_\nu\rangle+ L_3
\left\langle (u_\mu u^\mu)^2\right\rangle +L_4 \langle u^\mu u_\mu\rangle
\langle\chi_+\rangle \no &  & +L_5\langle u^\mu u^{}_\mu \chi^{}_+\rangle + L_6 \langle
\chi^{}_+\rangle^2 + L_7
\langle\chi^{}_-\rangle^2+\frac{L_8}{2}\left\langle\chi_+^2+\chi_-^2\right\rangle
- iL_9 \left\langle f_+^{\mu\nu}u^{}_\mu u^{}_\nu\right\rangle \no &  &
+\frac{L_{10}}{4}\left\langle f_+^2-f_-^2\right\rangle +H_1 \left\langle
F_L^2+F_R^2\right\rangle + H_2 \langle \chi \chi^\dag \rangle,
\label{lagphi}
\eea
where $D^{\mu\nu\rho}=\{D^{}_\mu,
\{D^{}_\nu,D^{}_\rho\}\}$,$F_L^{\mu\nu}=\partial^\mu l^\nu-\partial^\nu l^\mu
-i[l^\mu,l^\nu]$, $F_R^{\mu\nu}=\partial^\mu r^\nu-\partial^\nu r^\mu -i[r^\mu,r^\nu]$, and $f_{\pm}^{\mu\nu}=uF_L^{\mu\nu}u^\dag\pm u^\dag F_R^{\mu\nu}u$.

\section{One-loop generating functional}\label{secfunc}

The generating functional up to $\order{p^3}$ consists of two
parts~\footnote{The chiral anomalous effective Wess--Zumino--Witten
action~\cite{Wess:1971yu,Witten:1983tx} is not taken into account here.}:
\bea
Z_3=Z_3^\text{tree}+Z_3^\text{one-loop}.
\eea
The tree-level part $Z_3^\text{tree}$ is given by the Lagrangians in
Eqs.~\eqref{lagphiP} and \eqref{lagphi}. The one-loop functional
$Z_3^\text{one-loop}$ can be calculated in the standard way using the background
field method. To calculate the one-loop functional, we perturb the fields $U(x)$
and $P(x)$ around the solutions of classical equations of motion $\bar{U}(x)$
and $\bar{P}(x)$ as
\bea
U=\bar{u} e^{-i\eta} \bar{u}, \qquad P=\bar{P}+h,
\eea
with $\eta=\eta^a \lambda^a~(a=1,\ldots,N^2-1)$. In the following, we will
neglect the bars over the classical field configurations for brevity.
Collecting the fluctuations in $\xi_A=(\frac{F_0}{\sqrt{2}}\eta^a,h_i)$, the
one-loop functional can be written as a Gaussian integral over the fluctuations,
\bea
e^{iZ^\text{one-loop}} & = & \int [d\xi] \exp{ \left\{ -i
\int d^4x\, \xi^{}_A \left(\mathbb{D}_\mu \mathbb{D}^\mu
+\sigma\right)^{AB}\xi_B^\dag \right\} } \no & = & \mathcal{N}\exp \left\{
-\frac{1}{2}\text{tr}~\text{log} \left(\mathbb{D}_\mu \mathbb{D}^\mu +
\sigma\right)\right\}, \label{oneloopfunct}
\eea
where $\mathcal{N}$ is a normalization constant, ``tr" indicates the trace over
all the spaces including the coordinate space and the $(N^2-1+N)$-dimension
space spanned in the basis of $\xi_A$.
The covariant derivative $\mathbb{D}_\mu^{AB}$ and the non-derivative term
$\sigma^{AB}$ are obtained in Ref.~\cite{Du:2016ntw}. For simplicity, we only
list the single-matter sectors in following:
\bea
\mathbb{D}_\mu^{AB} & = & \delta^{AB}\partial_\mu +\hat{\Gamma}_\mu^{AB}=\begin{pmatrix}
    d_\mu^{ab} & \frac{1}{4\sqrt{2}F_0}\big( P[u_\mu, \lambda^a]\big)_j  \\
    \frac{1}{4\sqrt{2}F_0}\big( [u_\mu,\lambda^b]P^\dagger\big)_i &  D_\mu^{ij}
\end{pmatrix},  \\
\sigma^{AB}& = & \begin{pmatrix}
\sigma_{11}^{ab} & \sigma_{12}^{aj} \\
\sigma_{21}^{ib} & \sigma_{22}^{ij}
\end{pmatrix},
\eea
where
\bea
d_\mu^{ab} & = & \delta^{ab}\partial_\mu - \frac{1}{2} \tr{
[\lambda^a,\lambda^b]\Gamma_\mu} - \frac{1}{8F_0^2}\left(D_\mu P[\lambda^a,
\lambda^b]P^\dagger- P [\lambda^a,\lambda^b]D_\mu
P^\dagger\right)\, ,  \no
\sigma_{11}^{ab} & = & -\frac{1}{8}\tr{u_\mu
\left[\lambda^a,[u^\mu,\lambda^b]\right]} + \frac{1}{16}\tr{\left\{
\lambda^a,\{\chi_+,\lambda^b\}\right\}}
+\frac{3}{32F_0^2}P [u_\mu,\lambda^a][u^\mu,\lambda^b]P^\dagger ,
\nonumber\\
\sigma_{12}^{aj} & = & -\frac{1}{4\sqrt{2}F_0}\big( P[\nabla_\mu
u^\mu,\lambda^a]\big)_j-\frac{3}{4\sqrt{2}F_0}\big( D_\mu
P[u^\mu,\lambda^a]\big)_j\, , \nonumber\\
\sigma_{21}^{ib} & = & \frac{1}{4\sqrt{2}F_0} \left([\nabla_\mu
u^\mu,\lambda^b]P^\dagger\right)_i+\frac{3}{4\sqrt{2}F_0}\left([u^\mu,\lambda^b]
D_\mu P^\dagger\right)_i\, , \nonumber\\
\sigma_{22}^{ij} & = & m^2
\delta^{ij}-\frac{1}{32F_0^2}\left([u_\mu,\lambda^c]P^\dagger\right)_i\big(
P[u^\mu,\lambda^c]\big)_j\,.
\eea

To calculate the one-loop functional $Z^{\text{one-loop}}=\frac{i}{2}\text{tr log}
(\mathbb{D}_\mu \mathbb{D}^\mu +\sigma )$, we split
the differential operator $\mathcal{D}=(\mathbb{D}_\mu \mathbb{D}^\mu +\sigma )$
into the Klein-Gordon operator $\mathcal{D}_0$ for free fields and
the remaining interaction part $\delta_r$:
$\mathcal{D}^{AB}=\mathcal{D}^{AB}_0+\delta^{AB}_r$. The one-loop functional can
be calculated in an expansion of the interaction term:
\bea
Z^\text{one-loop}& = & \frac{i}{2}\text{tr log}(\mathcal{D}_0+\delta_r)=\frac{i}{2}\text{tr}
\big[ \text{log}\mathcal{D}_0+\text{log}(\mathbf{1}-\delta_r \Delta)\big], \no
& = & 
-\frac{i}{2}\text{tr}(\delta_r\Delta)-
\frac{i}{4}\text{tr}(\delta_r\Delta \delta_r \Delta)
-\frac{i}{6}\text{tr}(\delta_r\Delta \delta_r \Delta \delta_r \Delta)+\ldots, \label{oneloopfunctexpand}
\eea
where we omit the irrelevant constant term $\frac{i}{2}\text{tr
log}(\mathcal{D}_0)$, $\Delta$ is the inverse of $-\mathcal{D}_0$:
\bea
\Delta^{AB}(x-y)=\delta^{AB}\int \frac{d^dp}{(2\pi)^d}\frac{e^{-ip(x-y)}}{p^2-m_A^2+i\epsilon},
\eea
and the remainder $\delta_r$ is
\bea
\delta_r=\{ \hat{\Gamma}^\mu,\partial_\mu \}+\hat{\Gamma}^\mu \hat{\Gamma}_\mu+
\hat{\sigma}\,,
\eea
with $\hat{\sigma}^{AB}=\sigma^{AB}-m_A^2\delta^{AB}$. The first term
$\text{tr}(\delta_r\Delta)$ in Eq.~(\ref{oneloopfunctexpand}) is the set of all
tadpole graphs. The second term collects all the two-point-loop graphs, etc.
Since the external fields $v_\mu$ and $s$ ($a_\mu$ and $p$) correspond to terms
with an even (odd) number of boson fields, following the
counting scheme used in Refs.~\cite{Gasser:1983yg, Gasser:1984gg}, we count
$v_\mu$ and $s-\mathcal{M}$ as $\order{\phi^2}$, where $\phi$ here should be
understood as representing both the Goldstone boson and matter fields.
Thus, $\hat{\Gamma}_\mu$ and $\hat{\sigma}$ are of $\order{\phi^2}$. The the one-loop functional up to $\order{\phi^4}$,
i.e.
with at most four external meson fields, can be calculated as~\cite{Unterdorfer:2005au}
\bea
Z^\text{one-loop} & = & -\frac{i}{2}  \mbox{tr}\left[
\hat{\sigma(x)}\Delta(0)+\hat{\Gamma}^\mu(x)\hat{\Gamma}_\mu(x)\Delta(0)\right]
\nonumber \\
 & & -\frac{i}{4} \mbox{tr}\Big[ \left\{
  \hat{\Gamma}^\mu(x),\partial_\mu^x\right\} \Delta(x-y)
  \left\{\hat{\Gamma}^\nu(y), \partial_\nu^y\right\}\Delta(y-x) \nonumber\\
  & & +2 \left\{\hat{\Gamma}^\mu(x),\partial_\mu^x\right\}\Delta(x-y)
  \hat{\sigma}(y)\Delta(y-x)  \nonumber\\
  & &
  +2\hat{\Gamma}^\mu(x)\hat{\Gamma}_\mu(x)\Delta(x-y)\hat{\sigma}(y)\Delta(y-x)
  \Big] + \order{\phi^6}
  \nonumber  \\
  & = & -\frac{i}{2} \mbox{tr}\left[
 \hat{\sigma}(x)\Delta(0)+\hat{\Gamma}^\mu(x)\hat{\Gamma}_\mu(x)\Delta(0)
 \right]\nonumber \\
  & & -\frac{i}{4} \mbox{tr}\Big[
  \hat{\Gamma}^\mu(x)\partial_\mu^x\Delta(x-y)\hat{\Gamma}^\nu(y)\partial_\nu^y\Delta(y-x) \nonumber\\
  & & +\hat{\Gamma}^\mu(x)\partial_\nu^x \Delta(x-y)\hat{\Gamma}^\nu(y)\partial_\mu^y\Delta(y-x) \nonumber \\
  & & +\hat{\Gamma}^\mu(x)\partial_\mu^x\partial_\nu^x \Delta(x-y)\hat{\Gamma}^\nu(y)\Delta(y-x)\nonumber \\
  & & +\hat{\Gamma}^\mu(x)\Delta(x-y)\hat{\Gamma}^\nu(y)\partial_\mu^y\partial_\nu^y\Delta(y-x) \nonumber\\
  & & +2\hat{\Gamma}^\mu(x)\partial_\mu^x \Delta(x-y)\hat{\sigma}(y)\Delta(y-x) \nonumber\\
  & & +2
 \hat{\Gamma}^\mu(x)\Delta(x-y)\hat{\sigma}(y)\partial_\mu^y\Delta(y-x) \nonumber\\
  & & +\hat{\sigma}(x)\Delta(x-y)\hat{\sigma}(y)\Delta(y-x) \Big]+
 \order{\phi^6} .
 \nonumber
\eea

When the light quark masses are different, the free Klein-Gordon propagator is
not dia\-gonal in the cartesian basis spanned by $\lambda^1,\lambda^2,\ldots,
\lambda^{N^2-1}$. It is more convenient to use the physical basis such as
$\lambda^{\pi^+}=-\frac{1}{\sqrt{2}}(\lambda^1+i\lambda^2)$ and so on. More
explicitly, the one-loop functional to $\order{\phi^4}$ can be written
as~\cite{Gasser:1984gg}
\bea
Z^\text{one-loop}  & = &  \frac{i}{2} \sum_P \int d^4x  \Delta_P(0)\hat{\sigma}_{PP}(x)+\frac{i}{4}
\sum_{P,Q}\big[\Delta_P(0)+\Delta_Q(0) \big]\hat{\Gamma}_{\mu
PQ}(x)\hat{\Gamma}^\mu_{QP}(x) \no
 & & + \sum_{P,Q}\int d^4x d^4y \Big[
M_{\mu\nu}^{PQ}(x-y) \hat{\Gamma}^\mu_{PQ}(x)\hat{\Gamma}^\nu_{QP}(y) +
K_\mu^{PQ}(x-y)\hat{\Gamma}^\mu_{PQ}(x)\hat{\sigma}_{QP}(y)  \no
 & & + J^{PQ}(x-y) 
\hat{\sigma}_{PQ}(x)\hat{\sigma}_{QP}(y)
 \Big] + \order{\phi^6}, \label{eq:funct}
\eea
where
\bea
M_{\mu\mu}^{PQ}(z) & = & \frac{i}{4} \left( \partial_\mu \Delta_P \partial_\nu
\Delta_Q+\partial_\nu \Delta_P \partial_\mu \Delta_Q -\partial_{\mu\nu}
\Delta_P\Delta_Q-\Delta_P\partial_{\mu\nu}\Delta_Q \right),  \no
K_\mu^{PQ}(z)& = & \frac{i}{2}(\partial_\mu \Delta_P \Delta_Q-\Delta_P
\partial_\mu \Delta_Q), \no
J^{PQ}(z)& = & -\frac{i}{4}\Delta_P\Delta_Q,
\eea
with $\Delta_P(z)=\Delta(z,M_P^2)$ defined as the free propagator for a spinless
field of mass $M_P$ in $d$ dimensions.
The explicit expressions of the various kernels in terms of loop integrals are
listed in  Appendix.~\ref{app:ker}.

\section{Subtraction of the power counting breaking terms}
\label{secpcb}

In calculations using  dimensional regularization with the
$\overline{\text{MS}}$ scheme, any loop integral involving matter field
propagators contains terms starting from $\order{1}$, in contrast with the power
counting in Eq.~\eqref{pcr}. Thus, there are PCB terms which contribute at
orders lower than that required by Eq.~\eqref{pcr}, and they will be calculated
and subtracted in this section.

The loops involving only matter field propagators do not play any
dynamical role in the low-energy effective field theory, neither do they
introduce non-analyticity in the quark masses, and thus can be absorbed into a
redefinition of LECs. As a result, the explicit closed matter field loops are
not necessary to be included in the calculation.
Thus, the Goldstone boson part of the one-loop functional is identical to that
in the standard Goldstone boson ChPT, which can be found in
Refs.~\cite{Gasser:1984gg}.
The PCB terms of interest are from the loops containing both Goldstone boson and
matter field internal propagators. They correspond to the terms of
$M_{\mu\nu}^{PQ}$, $K_{\mu}^{PQ}$ and $J^{PQ}$ in Eq.~\eqref{eq:funct} with $P$
and $Q$ in different blocks (Goldstone bosons and matter fields).
Since we are only interested in the single matter field sector, which is the
sector relevant for processes with a single matter field in both initial and
final states and can be studied using the effective Lagragians in
Eq.~\eqref{lagphiP}, and the terms contributing at orders lower than that
required by Eq.~\eqref{pcr}, we only need the following terms of
$\hat\Gamma_\mu$ and $\hat\sigma$ for calculating the PCB part, which is of $\order{p^2}$, of the one-loop generating functional
\bea
\hat{\Gamma}_\mu^{\prime AB}& = & \frac{1}{4\sqrt{2}F_0}\begin{pmatrix}
0 & \left(P[u_\mu,\lambda^a]\right)_j \\
\left([u_\mu,\lambda^b]P^\dagger \right)_i & 0
\end{pmatrix}, \no
\hat{\sigma}^{\prime AB}& = &\frac{3}{4\sqrt{2}F_0}\begin{pmatrix}
0 & -\left(D^\mu P[u_\mu,\lambda^a]\right)_j \\
\left([u_\mu,\lambda^b]D^\mu P^\dagger\right)_i & 0
\end{pmatrix}.
\eea

Since the elements of $\Gamma^\prime_\mu$ and $\sigma^\prime$ are of
$\order{p}$, the PCB terms of loops only refer to the $\order{1}$ terms of the
loop integrals. As a result, they are independent of the internal Goldstone
boson masses.
Therefore, the relevant one-loop generating functional to $\order{\phi^4}$ can
be rewritten in the form
\bea
Z^{\prime \text{one-loop}} & = & \int d^4x d^4y \frac{d^4p}{(2\pi)^4}
e^{-ip(x-y)}\no & &
\times \bigg\{ \frac{1}{8} \left\langle \hat{\Gamma}^{\prime
\mu\nu}(x)\hat{\Gamma}^\prime_{\mu\nu}(y)\right\rangle
\left[B_0(p^2,m^2,0)+4B_1(p^2,m^2,0)+4B_{11}(p^2,m^2,0) \right] \no
& & - \left\langle \hat{\Gamma}^{\prime}_\mu (x)\hat{\Gamma}^{\prime
\mu}(y)\right\rangle \left[
B_{00}(p^2,m^2,0)+\frac{p^2}{4}\big(
B_0(p^2,m^2,0)+4B_1(p^2,m^2,0)+4B_{11}(p^2,m^2,0) \big) \right]   \no
& & + \left\langle\partial_\mu \tilde{\Gamma}^{\prime \mu} (x)
\hat{\sigma}(y)\right\rangle \left[ \frac{B_0(p^2,m^2,0)}{2}+B_1(p^2,m^2,0)
\right] +\frac{1}{4}B_0(p^2,m^2,0) \left\langle \hat{\sigma}^\prime
 (x)\hat{\sigma}^\prime (y)\right\rangle \bigg\}, \no
 & = & \int d^4x d^4y
 \frac{d^4p}{(2\pi)^4}e^{-ip(x-y)} \bigg\{ -\left\langle
 \hat{\Gamma}^{\prime}_\mu (x)\hat{\Gamma}^{\prime \mu}(y)\right\rangle
 B_{00}(p^2,m^2,0) \no
 & & - \frac14\left\langle \partial_\mu \Gamma^\prime_\nu \partial^\nu
 \Gamma^{\prime \mu}\right\rangle
 {\left[ B_0(p^2,m^2,0)+4B_1(p^2,m^2,0)+4B_{11}(p^2,m^2,0)\right]} \no
 & & +
 \left\langle\partial_\mu \tilde{\Gamma}^{\prime \mu} (x)
 \hat{\sigma}(y)\right\rangle \left[\frac{B_0(p^2,m^2,0)}{2}+B_1(p^2,m^2,0)
 \right] +\frac{1}{4}B_0(p^2,m^2,0) \left\langle \hat{\sigma}^\prime
 (x)\hat{\sigma}^\prime (y)\right\rangle \bigg\},   \label{eq:Zprime}
\eea
where we have defined $\hat \Gamma'_{\mu\nu} = \partial_\mu \hat \Gamma'_{\nu} -
\partial_\nu \hat \Gamma'_{\mu} + [\hat \Gamma'_{\mu}, \hat \Gamma'_{\nu}]$, and
\bea
\tilde{\Gamma}_\mu^{\prime AB}&= &\frac{1}{4\sqrt{2}F_0}\begin{pmatrix}
0 & \left(P[u_\mu,\lambda^a]\right)_j \\
-\left([u_\mu,\lambda^b]P^\dagger \right)_i & 0
\end{pmatrix} .
\eea
The expressions for the loop functions $B_0(s,m^2,M^2)$,
$B_1(s,m^2,M^2)$, $B_{00}(s,m^2,M^2)$ and $B_{11}(s,m^2,M^2)$ are given in
Appendix~\ref{app:ker}.

Since all the operators in Eq.~\eqref{eq:Zprime} are of $\order{p^2}$, we can
extract the PCB terms by keeping only the $\order{p^0}$ part of these loop
functions.
In the EOMS scheme, the PCB terms come from the leading chiral expansion of
one-loop functions $A(m^2)$ and $B_0(m^2,m^2,0)$, as shown in
Appendix~\ref{app:pcb}.
Applying the results of Appendix~\ref{app:pcb} and the equations of motion for
the classical background fields, it is easy to obtain the PCB
terms of interest:
\bea
Z_\text{one-loop}^{\text{PCB}} &= & \frac{1}{16\pi^2F_0^2}\int d^4x
\bigg\{
\frac{m^2}{144} \left[2-3 \mbox{log}\left(\frac{m^2}{\mu^2}\right)\right]
\left\langle P P^\dagger\right\rangle \left\langle u_\mu u^\mu\right\rangle
\nonumber \\
 & & +\frac{m^2 ~N}{144} \left[2-3
 \mbox{log}\left(\frac{m^2}{\mu^2}\right)\right]\left\langle P u_\mu u^\mu
 P^\dagger\right\rangle \nonumber \\
 & & +\frac{7}{72} \left[5-3
 \mbox{log}\left(\frac{m^2}{\mu^2}\right)\right] \left\langle D^\mu P D^\nu
 P^\dag\right\rangle \left\langle u_\mu u_\nu \right\rangle \nonumber \\
 & & +\frac{7N}{144} \left[5-3
 \mbox{log}\left(\frac{m^2}{\mu^2}\right)\right] \left\langle D^\mu
 P \left\{u_\mu,u_\nu\right\} D^\nu P^\dag \right\rangle \bigg\}. \label{pcb}
\eea

One can subtract these PCB terms in Eq.~\eqref{pcb} to get a consistent power
counting. Within the EOMS scheme, they are absorbed into the redefinition of the
LECs of $\order{p^2}$ as
\bea
\mathcal{L}_{\phi P}^{(2)}=\sum_{i=0}^{5} h_i \mathcal{O}_i= \sum_{i=0}^{5}
\left[ h_i^r(\mu) + h_i^0\lambda+ \frac{1}{16\pi^2 F_0^2}h_i^{\text{PCB}}
\right]\mathcal{O}_i\,,
\eea
where $\mathcal{O}_i$ represent local operators in the Lagrangian of
$\order{p^2}$, $\mu$ is the scale of  dimensional regularization,
$\lambda=\mu^{d-4}(4\pi)^{-d/2}/(d-4)$, $h_i^r(\mu)$ are the UV finite and
scale-dependent part of the LECs $h_i$, the coefficients $h_i^0$ of the UV
divergence $\sim \lambda$ have been calculated in Ref.~\cite{Du:2016ntw}, and
$h_i^\text{PCB}$ are the PCB parts.
From Eq.~\eqref{lagphiP} and Eq.~\eqref{pcb}, they can be easily read off as
\bea
h_0^{\text{PCB}}& = & 0, \quad
h_2^\text{PCB}=-m^2 \left(\frac{1}{72}-\frac{1}{48}\log
\frac{m^2}{\mu^2}\right), \quad
h_3^\text{PCB}=m^2\left(\frac{N}{72}-\frac{N}{48}\log \frac{m^2}{\mu^2}\right),
\no
h_1^{\text{PCB}}& = & 0,\quad h_4^{\text{PCB}} =-\frac{7}{72}\left(5-3\log
\frac{m^2}{\mu^2}\right), \qquad~\, h_5^\text{PCB}=\frac{7N}{144}\left(5-3\log
\frac{m^2}{\mu^2}\right).
\eea
Note that we have dropped the tadpole loops with matter field propagators
completely. For $N=3$, the expressions of $h_i^\text{PCB}$ agree with those
found in an explicit calculation of the charmed-meson--Goldstone-boson
scattering amplitudes~\cite{Yao:2015qia}.

\section{Summary}\label{summary}

In this paper, we have given the explicit generating functional for Green
functions of at most four external fields in a chiral effective
field theory for a single matter field up to $\order{p^3}$. In the case we
considered, the matter fields are spinless and in the fundamental representation
of $\text{SU}(N)$.
We have derived the power counting breaking terms in the one-loop generating
functional up to $\order{p^3}$ using dimensional regularization. In the EOMS
scheme, they are subtracted from the loop integrals and absorbed into a
redefinition of $\order{p^2}$ LECs.
The framework can be used for any theories with spontaneous
symmetry breaking of \sulr~to \suv~with spinless matter fields in fundamental
representation. Examples in QCD of the matter fields are ground state
pseudoscalar mesons except for the pions such as kaons and heavy mesons.

\section*{Acknowledgements}

UGM acknowledges the warm hospitality of the ITP of CAS where part of this work was done.
This work is supported in part by DFG and NSFC through funds provided to the
Sino-German CRC 110 ``Symmetries and the Emergence of Structure in QCD" (NSFC
Grant No.~11621131001), by the Thousand Talents Plan for Young Professionals, by
the Chinese Academy of Sciences (CAS) (Grant No.~QYZDB-SSW-SYS013), and by the
CAS President's International Fellowship Initiative (PIFI) (Grant
No.~2015VMA076).

\bigskip

\begin{appendix}

\section{Kernels} \label{app:ker}

The kernels $\mathcal{K}_a^{PQ}(x-y)$ including $M^{PQ}_{\mu\nu}
(x-y),~K^{PQ}_\mu (x-y)$ and $J^{PQ}(x-y)$ in Eq.~\eqref{eq:funct} have the
form
\bea
\mathcal{K}_a^{PQ}(x-y)=\int \frac{d^4p}{(2\pi)^2}e^{-ip(x-y)}\mathcal{K}_a^{PQ}(p),
\eea
with
\bea
M_{\mu\nu}^{PQ}(p)& = & -g_{\mu\nu} B_{00}(p^2,m_P^2,m_Q^2) \no
& & -\frac14 p_\mu p_\nu
\left[B_0(p^2,m_P^2,m_Q^2)+4B_1(p^2,m_P^2,m_Q^2)+
4B_{11}(p^2,m_P^2,m_Q^2)\right],\no
K_\mu^{PQ}(p)& = & ip_\mu \left[
\frac{1}{2}B_0(p^2,m_P^2,m_Q^2)+B_1(p^2,m_P^2,m_Q^2)\right], \no
J^{PQ}(p)& = & \frac{1}{4}B_0(p^2,m_P^2,m_Q^2),
\eea
where the loop functions $A$, $B_0$, $B_1$, $B_{00}$ and $B_{11}$ are defined through:
\bea
A(m^2) & = & \frac{\mu^{4-d}}{i}\int
\frac{d^dk}{(2\pi)^d}\frac{1}{k^2-m^2+i\epsilon},\\
B_0(p^2,m^2,M^2)& = & \frac{\mu^{4-d}}{i}\int
\frac{d^dk}{(2\pi)^d}\frac{1}{(k^2-m^2+i\epsilon)[(k+p)^2-M^2+i\epsilon]},\\
p^\mu B_1(p^2,m^2,M^2)& = & \frac{\mu^{4-d}}{i}\int
\frac{d^dk}{(2\pi)^d}\frac{k^\mu}{(k^2-m^2+i\epsilon)[(k+p)^2-M^2+i\epsilon]},\\
g^{\mu\nu}B_{00}(p^2,m^2,M^2) & + & p^\mu p^\nu B_{11}(p^2,m^2,M^2) \\
 & = &  \frac{\mu^{4-d}}{i}\int \frac{d^dk}{(2\pi)^d} \frac{k^\mu
 k^\nu}{(k^2-m^2+i\epsilon)[(k+p)^2-M^2+i\epsilon]}.
\eea
The expressions for these loop functions in dimensional regularization are
\bea
A_0(m^2) & = & \frac{m^2}{16\pi^2}\left( R +\log \frac{m^2}{\mu^2} \right) ,
\no B_0(p^2,m^2,M^2)& = & \frac{1}{16\pi^2}\bigg[-R+1-\log \frac{M^2}{\mu^2}
\no &  & +\frac{\Delta+p^2}{2p^2}
\log \frac{M^2}{m^2}+\frac{p^2-(m-M)^2}{p^2}\rho (p^2) \log
\frac{\rho(p^2)-1}{\rho (p^2)+1} \bigg],\no
B_1(p^2,m^2,M^2)& = &
\frac{1}{2p^2}\left[A_0(m^2)-A_0(M^2)-(p^2+\Delta)B_0(p^2,m^2,M^2)\right], \no
B_{00}(p^2,m^2,M^2)& = & -\frac{1}{288\pi^2}(p^2-3\Sigma)+\frac{1}{12p^2}\Big\{
(p^2+\Delta)A_0(m^2)+(p^2-\Delta)A_0(M^2) \no & & + \left[4p^2
m^2-(p^2+\Delta)^2\right]B_0(p^2,m^2,M^2)\Big\} \no
B_{11}(p^2,m^2,M^2)& =& \frac{1}{288\pi^2
p^2}(p^2-3\Sigma)+
\frac{1}{3p^4}\Big\{ -(p^2+\Delta)A_0(m^2)+(2p^2+\Delta)A_0(M^2) \no &  &- \left[p^2
m^2-(p^2+\Delta)^2\right]B_0(p^2,m^2,M^2)\Big\}, \label{eq:ab}
\eea
where we have defined $R=\frac{2}{d-4}+\gamma_E-1-\log 4\pi$, with $\gamma_E$ the Euler constant,
$\Delta = m^2-M^2 $, $\Sigma =m^2+M^2$, and
$$\rho=\sqrt{\frac{p^2-(m+M)^2}{p^2-(m-M)^2}}\,.$$

\section{Infrared regular parts of loop integrals} \label{app:pcb}

Using the method proposed in Refs.~\cite{Becher:1999he,Schindler:2003je}, we
derive the infrared regular parts of loop integrals to $\order{p^2}$. Only the
leading order, i.e. $\order{p^0}$, part is used to extract the PCB terms. The
closed matter loop $A_0(m^2)$ is infrared regular, and the regular part of the
loop integral $B_0(p^2,m^2,M^2)$ can be expanded as~\cite{Schindler:2003je}
\bea
B_0^\text{reg.}(p^2,m^2,M^2 )& = & \frac{\Gamma(2-d/2)}{(4\pi)^{d/2}(d-3)}
\left(\frac{m}{\mu}\right)^{d-4}\bigg[ 1- \frac{p^2-m^2}{2m^2}
+\frac{(d-6)(p^2-m^2)^2}{4m^4(d-5)} \no
&  & +\frac{(d-3)M^2}{2m^2(d-5)}+\cdots \bigg].
\eea
More explicitly, using Eq.~\eqref{eq:ab}, the infrared regular PCB parts of the loop
functions are
\bea
A_0^\text{PCB}(m^2)& = & -\frac{m^2}{16\pi^2}\log \frac{m^2}{\mu^2}, \no
B_0^\text{PCB}(p^2,m^2,0)& = & \frac{1}{16\pi^2}\left(1-\log
\frac{m^2}{\mu^2} \right), \no
B_1^\text{PCB}(p^2,m^2,0)& = &
-\frac{1}{16\pi^2}\left(1- \frac{1}{2}\log \frac{m^2}{\mu^2}\right), \no
B_{00}^\text{PCB}(p^2,m^2,0)& = & \frac{m^2}{288\pi^2}\left( 2-3\log
\frac{m^2}{\mu^2} \right), \no B_{11}^\text{PCB}(p^2,m^2,0)& = &
\frac{1}{144\pi^2}\left(8-3\log \frac{m^2}{\mu^2}\right).
\eea

\end{appendix}


\end{document}